\documentclass[11pt]{article}

\usepackage[english]{babel}

\usepackage[letterpaper,top=2cm,bottom=2cm,left=3cm,right=3cm,marginparwidth=1.75cm]{geometry}

\usepackage{cite,enumitem,subfigure}
\usepackage{amssymb}
\usepackage{amsmath,bm}
\usepackage{siunitx}
\PassOptionsToPackage{hyphens}{url}\usepackage{hyperref}
\usepackage{cleveref}
\usepackage[utf8]{inputenc}
\usepackage[right]{lineno}
\usepackage{csquotes}
\usepackage{booktabs}
\usepackage{longtable}
\usepackage{adjustbox}
\usepackage{array}
\usepackage{url}
\usepackage{appendix}
\usepackage{titlesec}
\usepackage{authblk}
\usepackage{xcolor} 

\usepackage{bbding}

\usepackage{booktabs,array,multirow}

\def\vf{{\bm{f}}}
\def\vg{{\bm{g}}}
\def\vh{{\bm{h}}}

\def\vq{{\bm{q}}}
\def\vr{{\bm{r}}}

\def\vv{{\bm{v}}}

\def\vx{{\bm{x}}}

\def\mD{{\bm{D}}}
\def\mE{{\bm{E}}}
\def\mF{{\bm{F}}}
\def\mG{{\bm{G}}}
\def\mH{{\bm{H}}}

\def\mL{{\bm{L}}}

\def\mP{{\bm{P}}}
\def\mQ{{\bm{Q}}}

\def\mS{{\bm{S}}}

\def\mW{{\bm{W}}}
\def\mX{{\bm{X}}}

\def\mZ{{\bm{Z}}}

\crefformat{figure}{#2Figure~#1#3}
\Crefformat{figure}{#2Figure~#1#3}
\crefformat{table}{#2Table~#1#3}
\Crefformat{table}{#2Table~#1#3}
\crefformat{section}{#2Section~#1#3}
\Crefformat{section}{#2Section~#1#3}

\author{Shuai Huang, Xuan Kan, James J. Lah, and Deqiang Qiu\thanks{Shuai Huang is with the Department of Electrical and Computer Engineering, Auburn University, AL 36830 USA (email: shuai.huang@auburn.edu). Xuan Kan is with the Department of Computer Science, Emory University, GA 30322 USA (email: xuan.kan@emory.edu). James J. Lah is with the Department of Neurology, Emory University, Atlanta, GA 30322, USA (jlah@emory.edu). Deqiang Qiu is with the Department of Radiology and Imaging Sciences, Emory University, GA 30322 USA (email: deqiang.qiu@emory.edu). }}

\title{\bfseries Atlas-free Brain Network Transformer}

\begin{document}
\maketitle

\begin{abstract}
Current atlas-based approaches to brain network analysis rely heavily on standardized anatomical or connectivity-driven brain atlases. However, these fixed atlases often introduce significant limitations, such as spatial misalignment across individuals, functional heterogeneity within predefined regions, and atlas-selection biases, collectively undermining the reliability and interpretability of the derived brain networks. To address these challenges, we propose a novel atlas-free brain network transformer (atlas-free BNT) that leverages individualized brain parcellations derived directly from subject-specific resting-state fMRI data. Our approach computes ROI-to-voxel connectivity features in a standardized voxel-based feature space, which are subsequently processed using the BNT architecture to produce comparable subject-level embeddings. Experimental evaluations on sex classification and brain-connectome age prediction tasks demonstrate that our atlas-free BNT consistently outperforms state-of-the-art atlas-based methods, including elastic net, BrainGNN, Graphormer and the original BNT. Our atlas-free approach significantly improves the precision, robustness, and generalizability of brain network analyses. This advancement holds great potential to enhance neuroimaging biomarkers and clinical diagnostic tools for personalized precision medicine. Reproducible code is available at \url{https://github.com/shuai-huang/atlas_free_bnt}
\end{abstract}

\section{Introduction}
\label{sec:introduction}

Functional Magnetic Resonance Imaging (fMRI) has become an indispensable tool in neuroscience research, it enables the non-invasive mapping of brain function by tracking blood oxygen level-dependent (BOLD) signals over time. A critical step in many resting state-fMRI data analysis pipelines involves parcellating the brain into regions of interest (ROIs), which serve as the nodes for subsequent functional connectivity and brain network analyses. The standard workflow typically begins by spatially normalizing each subject’s brain images to a common coordinate system (most commonly the Montreal Neurological Institute (MNI) template) followed by segmenting the brain into distinct ROIs according to a ``\emph{predefined}'' atlas. Such atlases are generally constructed using either prior anatomical knowledge or data-driven connectivity patterns among voxels. Anatomy-based atlases, such as the AAL-atlas \cite{tzourio2002automated} and Desikan-Killiany atlas \cite{desikan2006automated}, define regions based on structural landmarks such as gyri, sulci, or histological boundaries. In contrast, connectivity-based atlases, such as the Craddock atlas \cite{craddock2012whole} and the Schaefer parcellation derived from the human connectome project \cite{schaefer2018local}, rely on fMRI time-series data to infer inter-voxel functional connectivities, grouping together voxels with similar temporal profiles through clustering techniques.

\begin{figure*}[tbp]
\centering
\subfigure[ROI misalignment]{
\label{fig:misalignment}
\includegraphics[width=0.45\textwidth]{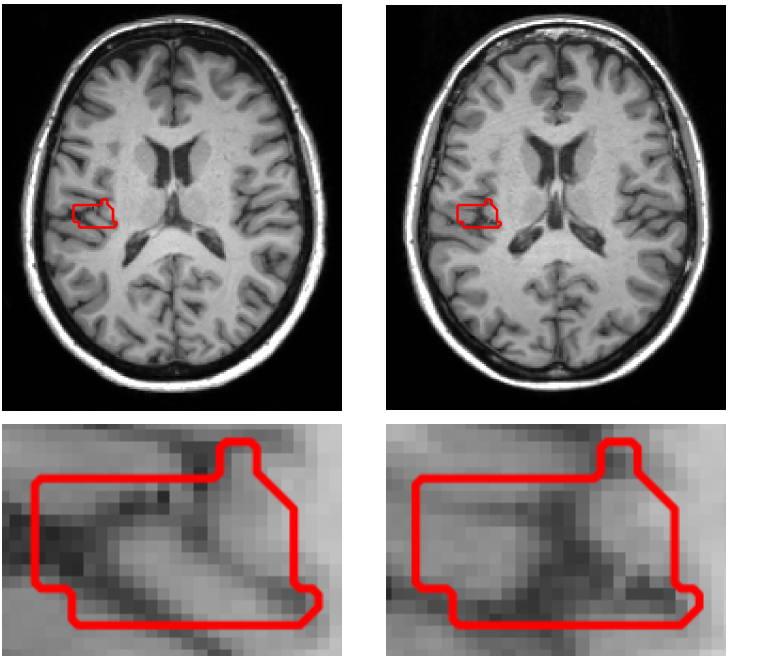}}
\subfigure[functional heterogeneity]{
\label{fig:heterogeneity}
\includegraphics[width=0.45\textwidth]{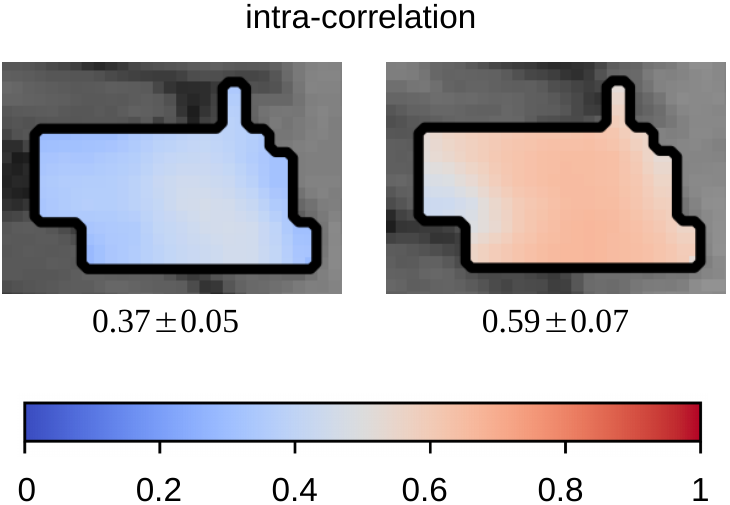}}
\caption{The Shen-368 brain atlas applied to standardized T1-weighted images in MNI space from two different subjects. Taking ROI $\#$52 as an example, both anatomical misalignment and functional heterogeneity arise due to inter-individual variability. (a) A zoomed-in view highlights the anatomical misalignment of the ROI between subjects. (b) Within this ROI, intra-correlations between the voxel-wise and mean BOLD time series are low, indicating substantial internal functional heterogeneity.}
\label{fig:fixed_atlas_issues}
\end{figure*}

Despite their widespread use, atlas-based analyses are vulnerable to atlas-selection bias: results can change materially depending on which parcellation is chosen. Differences in parcel number and size, boundary placement, and whether the atlas is anatomy- or connectivity-driven alter how BOLD signals are averaged, which in turn changes ROI community structure and downstream graph metrics. Consequently, group effects and model performance (e.g., classification accuracy) can vary across atlases even on the same dataset, undermining reproducibility and complicating cross-study comparisons. Moreover, most conventional atlases are derived from group-level datasets. This can lead to anatomical and/or functional misalignments when a common atlas is applied to individual subjects (see Fig. \ref{fig:misalignment}). Such misalignments may obscure meaningful subject-specific connectivity patterns, limiting the effectiveness of atlas-based approaches for personalized analysis. Another key limitation lies in the common assumption of functional homogeneity within each ROI. In practice, many atlas-defined regions contain considerable internal heterogeneity (see Fig. \ref{fig:heterogeneity}), which can dilute or distort connectivity estimates and reduce statistical sensitivity.

To address the limitations of traditional atlas-based analyses, we propose an atlas-free approach for constructing brain networks directly from each subject’s resting-state fMRI (rs-fMRI) data. Rather than relying on predefined, group-level atlases, our method derives individualized functional parcellations by clustering voxels that exhibit coherent intrinsic connectivity. This personalized parcellation strategy reduces spatial misalignment and improves the functional homogeneity of each ROI, enabling more accurate characterization of subject-specific brain connectivity. However, since each subject’s brain network is uniquely defined, direct cross-subject comparisons at the network level become challenging. To overcome this, we introduce the ``\emph{Atlas-free Brain Network Transformer}'' that is designed to extract robust, standardized, high-dimensional features from each subject’s unique functional connectivity map. These learned representations support group-level analyses such as classification and regression while preserving the precision of individualized parcellation. Our main contributions are summarized as follows:
\begin{itemize}
    \item Individualized Functional Parcellation: We present a data-driven method for generating individualized functionally coherent brain regions directly from rs-fMRI data, eliminating reliance on external brain atlases and addressing issues of misalignment and within-ROI heterogeneity.
    \item Atlas-free Brain Network Transformer: We develop a atlas-free transformer architecture that learns standardized feature representations from subject-specific brain networks, enabling consistent and meaningful group-level comparisons across subjects.
    \item Unified Analytical Framework:  We establish a unified approach that integrates personalized connectivity analysis with downstream group-level statistical inference, enhancing sensitivity and interpretability in clinical and cognitive neuroscience research.
\end{itemize}

\section{Related Work}
\subsection{Individualized Brain Parcellation}
Recent work on individualized brain parcellation shows that tailoring regions of interest (ROIs) to each subject’s intrinsic functional connectivity markedly improves both anatomical alignment and functional homogeneity, thereby overcoming key limitations of fixed, group-level atlases. At scale, Hermosillo et al. \cite{hermosillo2024precision} released the MIDB Precision Brain Atlas, generating subject-specific network maps across multiple cohorts and showing that these maps boost both test–retest reliability and brain–behavior correlations. Gordon et al. \cite{gordon2017precision} established that individual functional connectomes better captures idiosyncratic functional topography than population-based templates. Using a hybrid strategy, Wang et al. \cite{wang2015parcellating} introduced an iterative, population-guided refinement that produces subject-specific networks with markedly improved within-subject consistency across sessions. Recent machine-learning advances have further streamlined personalization: Qiu et al. \cite{qiu2022unrevealing} applied graph neural networks (GNNs) to model individualized cortical parcels, achieving high test–retest consistency while maintaining subject-specific distinctions. while Hu et al. \cite{hu2024consecutive}’s contrastive CC-SUnet yields reliable individualized parcellations from short scans, a critical step for clinical feasibility. Complementary evidence from Molloy \& Osher \cite{molloy2023personalized} shows that resting-state–derived personalized parcels outperform probabilistic atlases in predicting task-evoked functional ROIs for vision, language, motor, and working-memory systems. Finally, Li et al. \cite{li2024individual} provides a survey of recent advances, highlighting machine learning’s growing role in achieving scalable, reliable individualized parcellations.

\subsection{Deep Learning Approaches for Brain Network Analysis}
Deep-learning models such as graph neural network (GNN) \cite{kipf2017semisupervised,xu2019how,velickovic2017graph} and transformer \cite{vaswani2017attention,dwivedi2021generalization,kreuzer2021rethinking,ying2021transformers} have recently become prominent tools for brain network analysis. Early progress came from GNNs applied to functional-connectivity (FC) graphs. BrainNetCNN \cite{kawahara2017brainnetcnn} introduced convolutional filters that incorporate the topological structure of brain networks and outperformed fully-connected neural networks on demographic prediction. A spectral graph convolutional network (GCN) by Arslan et al. \cite{arslan2018graph} subsequently identified ROIs whose connectivity patterns correlate with different brain functions. Ktena et al. \cite{ktena2018metric} advanced this line by training a Siamese GCN to learn graph-similarity metrics for autism-spectrum-disorder diagnosis. To improve interpretability, BrainGNN \cite{li2021braingnn} introduced ROI-selection pooling that automatically highlights the most salient regions while maintaining strong predictive performance.

Transformer models, known for their ability to capture global context and long-range dependencies, have been increasingly adopted for brain network analysis.  The brain network transformer (BNT), proposed by Kan et al. \cite{kan2022brain}, applies a graph-transformer framework in which ROI-to-ROI correlation coefficients derived from rs-fMRI data are used as node features, and the connectivity patterns among ROIs are learned in a task-specific manner through self-attention. Unlike conventional transformer architectures, BNT incorporates a customized readout function that aggregates information from node embeddings to generate representations for downstream analysis tasks. To identify functional modules within the brain network, Dai et al. \cite{dai2023transformer} later introduced a transformer-based approach that jointly performs hierarchical clustering and classification. Further developments by Kan et al. \cite{kan2023dynamic} and Kim et al. \cite{kim2023swift} extended transformer architectures to capture the dynamic nature of functional connectivity over time.

Seeking to avoid the limitations of fixed brain atlases such as spatial misalignment and intra-ROI heterogeneity, some studies have explored the direct analysis of preprocessed rs-fMRI volumes instead of relying on ROI-based functional connectivity. For example, Malkiel et al. \cite{malkiel2022self} trained a multi-layer transformer to reconstruct fMRI data and fine-tuned the model for specific downstream tasks. Similarly, Sarraf et al. \cite{sarraf2023ovitad} applied a vision transformer to both rs-fMRI and structural MRI data for predicting Alzheimer’s disease progression. However, even after preprocessing, voxel-wise rs-fMRI data typically consists of over $\mathcal{O}(10^5)$ noisy, spatially correlated time series per subject, posing significant computational challenges and increasing the risk of overfitting due to low signal-to-noise (SNR) ratio. Parcellating the brain into functionally coherent ROIs and averaging signals within each region greatly reduces data dimensionality and improves SNR. Moreover, analyzing functional connectivity between ROIs produces interpretable brain-network graphs that yield more meaningful insights into the brain’s functional organization.

\section{Proposed Approach}
\subsection{Brain Parcellation via Clustering}
\label{subsec:clustering}
To ensure functional coherence within each ROI, we perform individualized brain parcellation by applying agglomerative or spectral clustering to group voxels with high pairwise functional connectivity together. Let $\vv_i,\vv_j\in\mathbb{R}^L$ denote two BOLD time series of length $L$. We measure the functional connectivity using the Pearson correlation coefficient $\rho_{ij}$:
\begin{align}
\rho_{ij}=\rho(\vv_i,\vv_j) = \frac{\textnormal{cov}(\vv_i,\vv_j)}{\sqrt{\textnormal{Var}(\vv_i)\cdot\textnormal{Var}(\vv_j)}},
\end{align}
where $\textnormal{cov}(\cdot)$ is the covariance, $\textnormal{Var}(\cdot)$ is the variance, $\rho_{ij}\in[-1,1]$ and measures the linear correlation between two time series.

\begin{enumerate}[label=\arabic*)]
\item {\bfseries Spatially-constrained Agglomerative Clustering}

With each voxel initially forming its own cluster, agglomerative clustering works in a bottom-up manner by iteratively merging the two most correlated clusters at each step until a stopping criterion is met. Let $\vr_i,\vr_j\in\mathbb{R}^L$ denote the centroids of two clusters $c_i$ and $c_j$ respectively, i.e. the averaged BOLD time series within each cluster. Following the centroid linkage strategy, we measure the similarity between clusters using the Pearson correlation coefficient $\rho(\vr_i,\vr_j)$. The merging process stops when either a predefined number of clusters is reached or the maximum correlation between any two clusters falls below a prespecified threshold $\nu$.

The standard agglomerative clustering algorithm has a computational complexity of $\mathcal{O}(N^3)$, where $N$ is the number of voxels. Although optimized implementations can reduce this to $\mathcal{O}(N^2)$, it remains computationally prohibitive for whole-brain scans, where $N\sim \mathcal{O}(10^5)$. However, functionally coherent voxels tend to be spatially contiguous, forming compact regions in the brain. Leveraging this observation, we constrain agglomerative clustering to merge only spatially adjacent voxels or clusters. This spatial constraint significantly reduces the computational burden to $\mathcal{O}(N\log N)$, making the method practical for large-scale, whole-brain parcellation without substantial loss of performance.

\item {\bfseries Spectral Clustering}

Spectral clustering partitions the voxels by leveraging the eigenspectrum of a similarity graph rather than relying on purely local, greedy merges. Starting from a similarity matrix $\mS$, i.e. the Pearson-correlation matrix, we construct a graph Laplacian $\mL=\mD-\mS$, where $\mD$ is the diagonal degree matrix with $D_{ii}=\sum_jS_{ij}$. The top $k$ eigenvectors of $\mL$ embed the voxels into a low-dimensional Euclidean space in which clusters become linearly separable. In this paper we use the normalized variant of the Laplacian matrix $\mL_\textnormal{norm}=\mD^{-\frac{1}{2}}\mL\mD^{-\frac{1}{2}}$ for a more balanced parcellation. The k-means clustering is then applied on the top $k$ eigenvectors of $\mL_\textnormal{norm}$ in this spectral space to obtain the final parcellation.

The complexity of spectral clustering is largely dominated by the construction of the similarity matrix, which requires $\mathcal{O}(N^2)$ operations and imposes substantial memory demands. To mitigate this, the similarity matrix $\mS$ can be sparsified by setting the entry $S_{ij}=0$ whenever$|S_{ij}|<\tau$, where $\tau$ is a predefined threshold. This reduces storage and accelerates downstream computation. Once the matrix is sparsified, the top $k$ eigenvectors can be efficiently computed using the Lanczos algorithm, whose cost scales linearly with the number of non-zero entries in $\mL_\textnormal{norm}$.

\end{enumerate}

\begin{figure}[tbp!]
    \centering
    \includegraphics[width=\columnwidth]{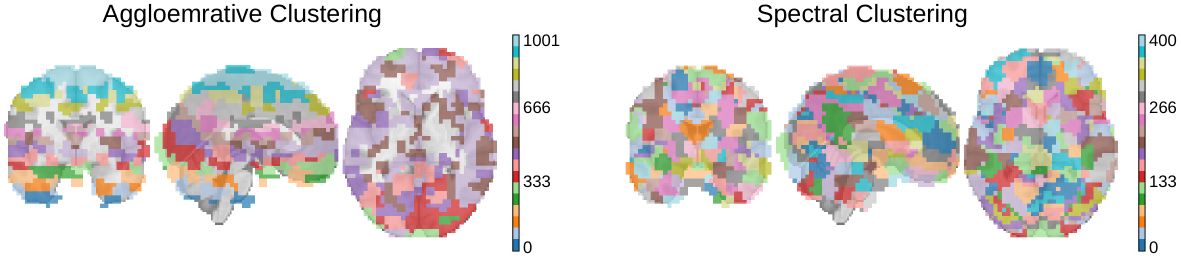}
    \caption{Individualized brain parcellation: the ROIs parcellated on a subject's brain using the agglomerative and spectral clustering methods.}
    \label{fig:ac_sc_atlases}
\end{figure}

As illustrated in Fig. \ref{fig:ac_sc_atlases}, both agglomerative and spectral clustering methods can partition the brain into a predetermined number of regions based on functional connectivity. Agglomerative clustering works hierarchically, relying primarily on local connectivity patterns; it iteratively merges spatially adjacent and highly correlated voxel clusters, thus ensuring that resulting clusters are spatially coherent and internally homogeneous. It also enables explicit control of cluster similarity through a threshold $\nu$ on the minimum allowed correlation between merged clusters. In contrast, spectral clustering leverages global connectivity information by embedding voxels into a low-dimensional space defined by the eigenvectors of the similarity graph Laplacian. This allows spectral clustering to identify functional communities that may span distant brain regions or exhibit complex, non-convex shapes, but it may be sensitive to noise and the choice of similarity threshold on $S_{ij}$. Together, these two methods offer complementary advantages for brain parcellation, balancing local coherence and global connectivity considerations.

\subsection{ROI-to-voxel Functional Connectivity}

After parcellating the brain into $R$ functionally homogeneous ROIs, we summarize the regional activity by averaging the BOLD signals within each ROI, obtaining mean time series $\{\vr_i\}_{i=1}^R$. However, individualized parcellations introduce variability in ROI definitions across subjects, making direct ROI-to-ROI connectivity comparisons infeasible for group-level brain network analysis. To address this challenge, we compute functional connectivity between each ROI and the voxels in the entire brain, thus ensuring a consistent connectivity feature space across subjects. Let $\vv_j$ denote the BOLD time series from the $j$-th brain voxel. The connectivity feature vector $\vf_i$ of the $i$-th ROI is
\begin{align}
    \vf_i=\left[\rho(\vr_i, \vv_1)\ \rho(\vr_i,\vv_2)\ \cdots\ \rho(\vr_i,\vv_D)\right]^\textnormal{T},
\end{align}
where $\rho(\cdot)$ is the Pearson correlation, $D$ is the number of non-zero voxels in the MNI space.

\begin{figure}[tbp!]
    \centering
    \includegraphics[width=0.8\columnwidth]{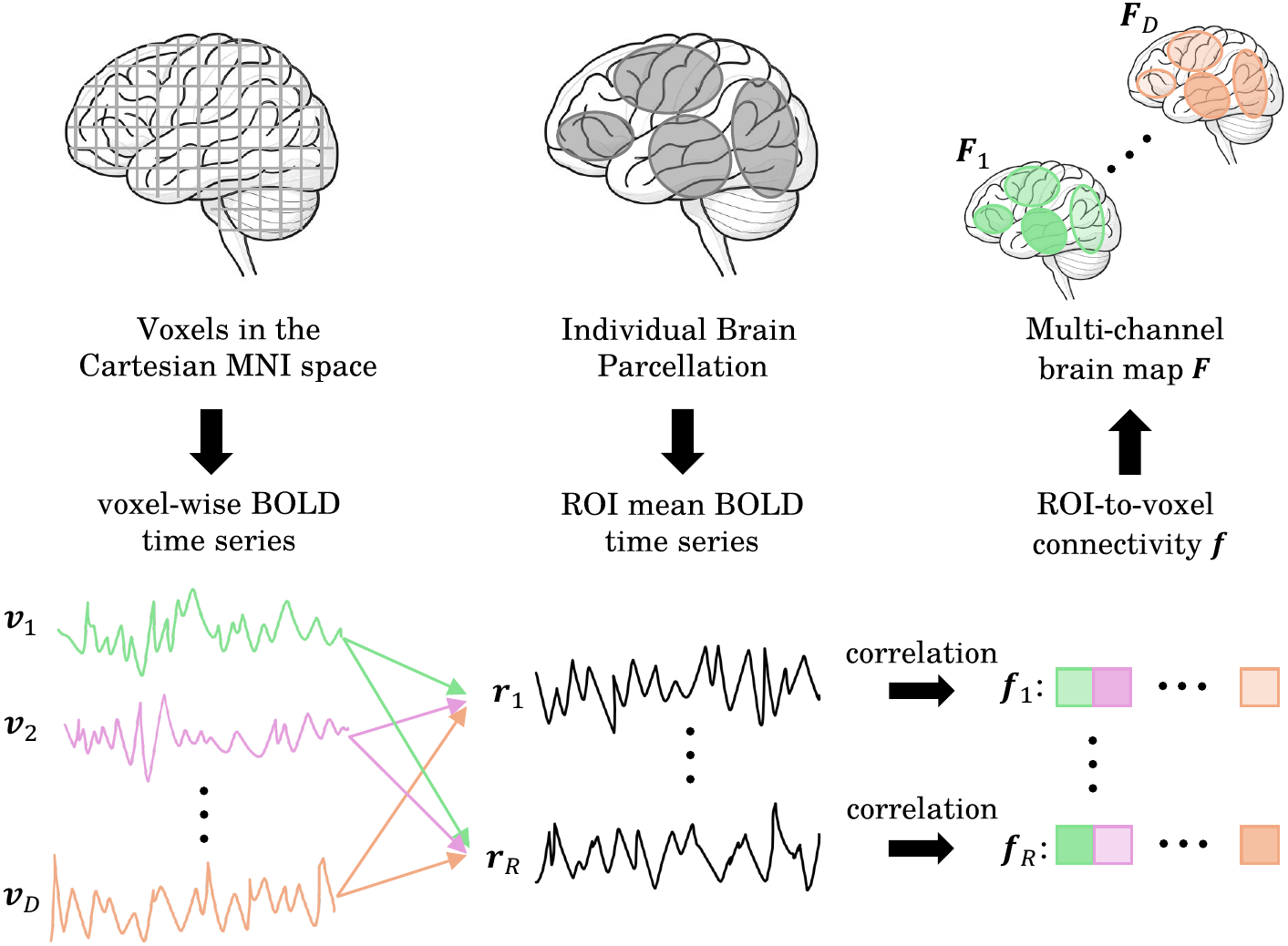}
    \caption{Each channel $\mF_j$ of the brain map $\mF=\{\mF_1,\cdots,\mF_D\}$ encodes the functional connectivity between the ROIs and a specific voxel $\vv_j$.}
    \label{fig:roi_to_voxel}
\end{figure}

As shown in Fig. \ref{fig:roi_to_voxel}, the ROI-to-voxel connectivities $\{\rho(\vr_i,\vv_j)\}_{i=1}^R$ that correspond to the same voxel $\vv_j$ are mapped back into the 3D space according to the ROIs' coordinates. In the standard MNI space of size $M_1\times M_2\times M_3$, this results in a multi-channel functional brain map $\mF\in\mathbb{R}^{M_1\times M_2\times M_3\times D}$:
\begin{align}
\mF=\left\{\mF_1,\, \cdots,\,\mF_j,\,\cdots,\,\mF_D\right\},
\end{align}
where each channel $\mF_j\in\mathbb{R}^{M_1\times M_2\times M_3}$ encodes the functional connectivities between the ROIs and the $j$-th voxel. 

\subsection{Atlas-free Brain Network Transformer}
Brain network analysis can then be performed based on the multi-channel brain map $\mF$. However, the number of voxels $D$ is typically on the order of $\mathcal{O}(10^5)$, resulting in a prohibitively large feature tensor $\mF$ and substantial computational overhead. To make the data tractable for transformer-based analysis, we need to first reduce its dimensionality. Specifically, we apply principal component analysis (PCA) to the connectivity vector $\vf_i$:
\begin{align}
    \vg_i=\mP\vf_i,
\end{align}
where $\vg_i\in\mathbb{R}^H$ is the resulting low-dimensional representation and $\mP\in\mathbb{R}^{H\times D}$ is the projection matrix learned from the development set.

\begin{figure}[tbp!]
    \centering
    \includegraphics[width=\columnwidth]{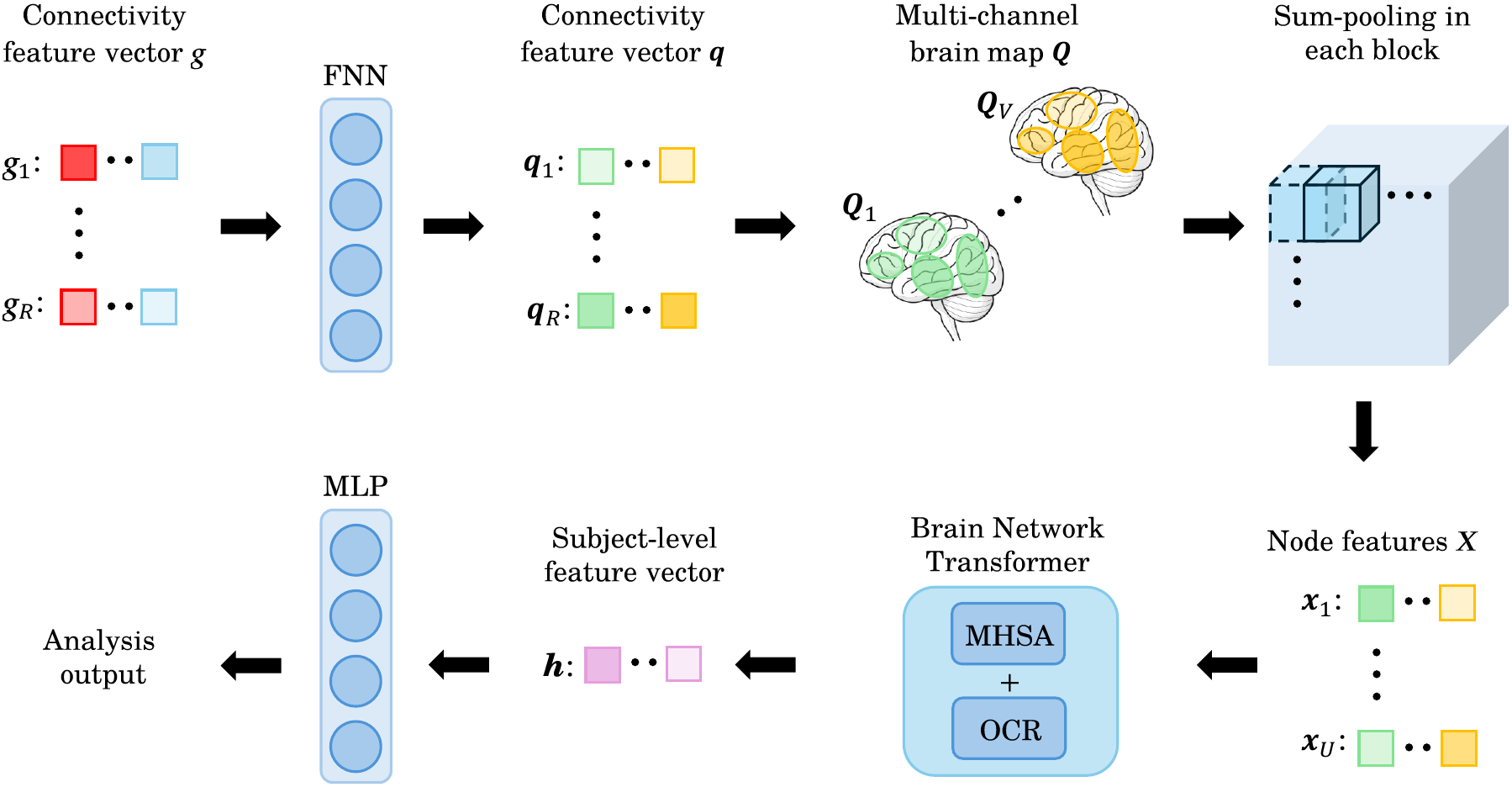}
    \caption{The proposed atlas-free brain network transformer.}
    \label{fig:af_bnt}
\end{figure}

As illustrated in Fig. \ref{fig:af_bnt}, the connectivity feature vector $\vg_i$ is first transformed by a feedforward neural network (FNN) into a new representation $\vq_i\in\mathbb{R}^V$. Following the same procedure used to construct $\mF$, we generate a new multi-channel brain map $\mQ\in\mathbb{R}^{M_1\times M_2\times M_3\times V}$ from $\{\vq_i\}$.
\begin{align}
\mQ=\left\{\mQ_1,\, \cdots,\,\mQ_j,\,\cdots,\,\mQ_V\right\}.
\end{align}

To extract standardized representations suitable for robust, cross-subject brain network analysis, we adopt the Brain Network Transformer (BNT) framework. BNT was originally developed to process atlas-based brain networks \cite{kan2022brain}, thus it cannot directly handle the voxel-level representation in $\mQ$. To bridge this gap, we partition the 3D MNI space into overlapping blocks of size $B\times B\times B$ with a stride $s$, and treat each spatial block as an input node to BNT. The node feature vector $\vx_i$ is then obtained by sum-pooling the voxel-wise features within the $i$-th block independently across all channels.

Let $\mX\in\mathbb{R}^{U\times V}$ represent the input node features, where $U$ is the the number of blocks/nodes and $V$ is the feature dimension. BNT employs a multi-head self-attention (MHSA) module consisting of $\mathcal{L}$ layers to obtain attention-enhanced node embeddings. Specifically, the output from the $l$-th layer, denoted $\mZ^l\in\mathbb{R}^{U\times V}$, is computed as follows:
\begin{align}
\mZ_m^{l} &=\textnormal{Softmax}\left(\frac{\mW^{l,m}_\mathcal{Q}\mZ^{l-1}\left(\mW^{l,m}_\mathcal{K}\mZ^{l-1}\right)^\textnormal{T}}{\sqrt{d^{l,m}_\mathcal{K}}}\right)\mW^{l,m}_\mathcal{V}\mZ^{l-1}\\
\mZ^l &= \textnormal{Concat}\left(\left.\mZ_m^{l}\right|_{m=1}^M\right)\mW^l_\mathcal{O},
\end{align}
where $\mZ_m^l$ is the output from the $m$-th attention head, $\{\mW^{l,m}_\mathcal{Q},\mW^{l,m}_\mathcal{K},\mW^{l,m}_\mathcal{V},\mW^l_\mathcal{O}\}$ are learnable model parameters, and $d^{l,m}_\mathcal{K}$ is the dimensionality of the key vectors for gradient stabilization. We can see that the query, key, and value matrices are all derived from the same input $\mZ^{l-1}$, i.e. the output from the previous layer. 


After the MHSA encoder, we apply the orthonormal clustering readout (OCR) proposed in \cite{kan2022brain} to aggregate the final-layer node embeddings $\mZ^{\mathcal{L}}$ into a compact, subject-level representation $\mH$. In OCR, the node embeddings are softly clustered into $K$ groups with learnable centroids $\mE\in\mathbb{R}^{K\times V}$. The soft assignment of node $i$ to cluster $k$ is computed via a softmax over inner products,
\begin{align}
\mP_{i,k}=\frac{\exp\left(\langle \mZ_i^{L}, \mE_k \rangle\right)}{\sum_{k'=1}^{K}\exp\left(\langle \mZ_i^{L}, \mE_{k'} \rangle\right)},
\end{align}
where $\langle\mZ_i^\mathcal{L},\mE_k\rangle$ denotes the inner product between the $i$-th node embedding and the $k$-th centroid. Given $\mP\in\mathbb{R}^{U\times K}$, the graph-level embedding is obtained by pooling node features within each cluster:
\begin{align}
    \mH=\mP^T\mZ
\end{align}

Finally, we vectorize $\mH\in\mathbb{R}^{K\times V}$ (i.e., $\vh=\textnormal{flatten}(\mH)$) and feed it into a multi-layer perceptron (MLP) for downstream tasks such as classification or regression.

\section{Experiments}
We evaluate the proposed atlas-free Brain Network Transformer (BNT) approach alongside leading atlas-based methods on two benchmark neuroimaging tasks: sex classification and brain-connectome age prediction. These experiments are designed to demonstrate the robustness and generalizability of the atlas-free framework and to showcase its ability to overcome the inherent limitations of fixed-atlas approaches.

\subsection{Experimental Setting}
\paragraph{Dataset.} We employ two distinct datasets for the aforementioned neuroimaging tasks:
\begin{itemize}
    \item \emph{Adolescent Brain Cognitive Development Study (ABCD).} The ABCD study is a 10-year longitudinal investigation tracking brain development in children starting at ages 9–10, with multiple follow-up sessions that include multimodal MRI scans and related behavioral assessments. To prevent data leakage into the test set and data redundancy in the training set, we randomly select only one session per subject for the sex classification task. After quality control, the final dataset comprises 6,738 subjects with biological sex labels, including 3,291 females (48.8$\%$) and 3,447 males (51.2$\%$).
    \item \emph{Emory Healthy Brain Study (EHBS).} The EHBS is a longitudinal study focusing on healthy older adults aged 50–90. Multimodal neuroimaging data and other health-related measures are collected, after obtaining written consent from the subjects and receiving approval from the Institutional Review Board (IRB) of Emory University. Given that subjects' biological ages change across follow-up sessions, multiple sessions from the same individual can be included in the brain-connectome age prediction dataset, provided all sessions from a given subject are consistently assigned to either the training or test set to prevent data leakage. Following quality control procedures, the dataset includes 2,255 sessions with recorded biological ages at the time of the MRI scan.
\end{itemize}

\paragraph{Evaluation.} To fully leverage the dataset, we conduct a 10-fold cross-validation. To mitigate the impact of random initialization, each fold's experiment is repeated 10 times. For the sex classification task, the final class label is determined by a majority vote among the 10 runs; in the event of a tie, we select the label corresponding to the model with the lowest training loss. For the brain-connectome age prediction task, we average the predicted ages from the 10 runs. The aggregated results across the 10 folds are reported to evaluate performance.

\begin{figure}[tbp!]
    \centering
    \includegraphics[width=\columnwidth]{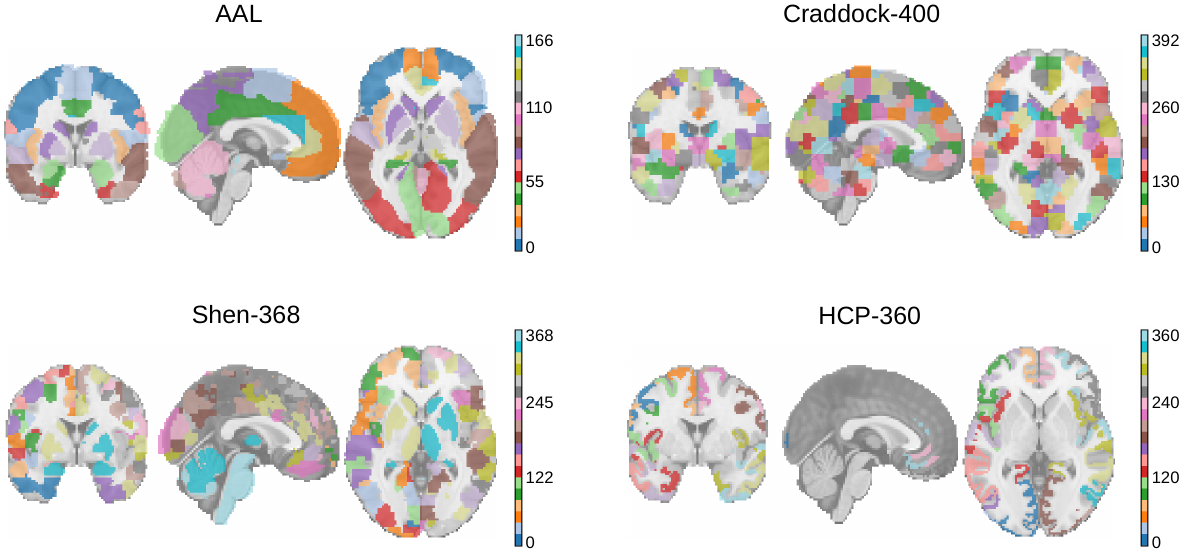}
    \caption{The atlas-based methods use four atlases to define ROIs in the brain: the AAL, Craddock-400, Shen-368 and HCP-360 atlases.}
    \label{fig:compare_atlases}
\end{figure}

\paragraph{Atlas-free BNT.} The raw resting-state fMRI measurements are processed using the CONN Toolbox \cite{nieto2020handbook, whitfield2012conn} to produce voxel-level BOLD signal time series in the standard 2mm MNI space. To reduce computational burden during individualized parcellation, we subsequently downsampled the time series to the 4mm MNI space. Based on the correlations between voxel time series, we perform individualized brain parcellation using the two clustering methods described in Section~\ref{subsec:clustering}. For spatially constrained agglomerative clustering, we set the merging correlation threshold to $\nu=0.8$, resulting in approximately 400 to 1200 clusters (ROIs) per subject. For spectral clustering, we sparsify the similarity matrix $\mS$ using a correlation cutoff threshold of $\tau=0.6$, and set the number of clusters to be $400$. Subsequently, ROI-to-voxel connectivity features are computed and organized into a multi-channel brain map $\mG$, forming the input feature tensor. For the atlas-free BNT analysis, we further partition the MNI volume into overlapping blocks of size $3\times 3\times 3$ voxels with a stride of $s=2$, yielding roughly 4600 input nodes. After extracting node features from each block, we pass them through a multi-head self-attention (MHSA) model with $L=1$ layer and $M=4$ attention heads, obtaining the final subject-level representation $\vh$. The atlas-free BNT model is trained using the Adam optimizer with a learning rate of $1e^{-6}$, a batch size of 16, and 50 epochs.

\paragraph{Baseline methods.} We compare our proposed atlas-free method with four established atlas-based baseline approaches: (1) the classical elastic net with $l_1$ and $l_2$ norm regularizations \cite{zou2005regularization}; (2) BrainGNN, which integrates ROI-selection pooling within a GNN architecture \cite{li2021braingnn}; (3) Graphormer, which extends the transformer framework for graph representation learning \cite{ying2021transformers} and (4) the original Brain Network Transformer (BNT), which leverages self-attention to learn connectivity patterns and the orthonormal clustering readout (OCR) function to aggregate node embeddings \cite{kan2022brain}. Hyperparameters for these baseline methods are tuned and fully detailed in Appendix~\ref{app:hyperparameter}. To mitigate atlas-selection bias inherent in traditional atlas-based analyses, we construct multiple brain networks using several widely adopted brain atlases. Specifically, we employ both anatomical and connectivity-driven atlases in the standard 2mm MNI space, including the anatomical AAL-atlas \cite{tzourio2002automated} and connectivity-based atlases such as Craddock-400 \cite{craddock2012whole}, Shen-368 \cite{shen2013groupwise}, and the HCP multimodal parcellation with 360 areas (HCP-360) \cite{glasser2016multi}. Fig. \ref{fig:compare_atlases} illustrates the ROIs defined by those four atlases.

\subsection{Sex Classification}

\begin{table}[tbp]
\caption{The performances of different approaches on the sex classification and brain-connectome age prediction tasks across 10-fold cross validation.}
\vspace{1em}
\label{tab:main_results}
\centering
\footnotesize
\resizebox{\columnwidth}{!}{
\begin{tabular}{llcccccccccccccccc}
\toprule
& &\multicolumn{4}{c}{\textbf{Sex classification}} &\textbf{Brain age prediction}\\ \cmidrule(lr){3-6} \cmidrule(lr){7-7}
\multirow{-2}{*}{Method} &\multirow{-2}{*}{Atlas} &Accuracy &Sensitivity &Specificity &AUROC & Mean Absolute Error  \\ \cmidrule(lr){1-2} \cmidrule(lr){3-6} \cmidrule(lr){7-7}
&AAL &78.17$\pm$1.63 &78.27$\pm$3.58 &77.92$\pm$2.49 &86.00$\pm$1.21 &5.61$\pm$0.27  \\
&Craddock-400 &86.78$\pm$1.05 &87.34$\pm$2.63 &86.22$\pm$2.84 &93.91$\pm$0.94 &4.86$\pm$0.24  \\
&Shen-368 &85.40$\pm$1.27 &86.28$\pm$2.91 &84.43$\pm$1.97 &92.95$\pm$1.02 &4.86$\pm$0.25  \\
\multirow{-4}{*}{Elastic net} &HCP-360 &84.59$\pm$0.90 &84.51$\pm$1.57 &84.69$\pm$2.82 &92.11$\pm$1.12 &4.90$\pm$0.20  \\
\cmidrule(lr){1-2} \cmidrule(lr){3-6} \cmidrule(lr){7-7}
&AAL &63.68$\pm$9.61 &79.21$\pm$21.0 &47.65$\pm$35.4 &69.60$\pm$13.9 &4.63$\pm$0.24  \\
&Craddock-400 &77.34$\pm$1.41 &76.81$\pm$2.71 &77.75$\pm$3.93 &84.62$\pm$1.15 &4.77$\pm$0.27  \\
&Shen-368 &74.83$\pm$1.66 &76.71$\pm$6.21 &72.87$\pm$5.29 &83.33$\pm$1.16 &4.88$\pm$0.40  \\
\multirow{-4}{*}{BrainGNN} &HCP-360 &69.50$\pm$10.32 &84.23$\pm$9.70 &54.31$\pm$29.9 &76.39$\pm$14.00 &4.79$\pm$0.26  \\
\cmidrule(lr){1-2} \cmidrule(lr){3-6} \cmidrule(lr){7-7}
&AAL &80.53$\pm$1.51 &83.73$\pm$3.04 &77.08$\pm$2.70 &88.25$\pm$1.20 &4.85$\pm$0.28  \\
&Craddock-400 &86.44$\pm$0.97 &89.08$\pm$2.04 &83.65$\pm$2.61 &93.99$\pm$0.67 &4.71$\pm$0.30  \\
&Shen-368 &85.77$\pm$1.49 &89.34$\pm$2.83 &81.99$\pm$3.30 &93.48$\pm$0.85 & 4.78$\pm$0.32 \\
\multirow{-4}{*}{Graphormer} &HCP-360 &84.51$\pm$1.02 &89.07$\pm$2.15 &79.75$\pm$2.93 &92.85$\pm$0.76 & 4.76$\pm$0.27 \\
\cmidrule(lr){1-2} \cmidrule(lr){3-6} \cmidrule(lr){7-7}
&AAL &79.10$\pm$1.42 &81.86$\pm$2.63 &76.18$\pm$2.60 &87.59$\pm$1.32 &4.50$\pm$0.27  \\
&Craddock-400 &87.93$\pm$1.08 &88.22$\pm$2.09 &87.65$\pm$2.18 &94.56$\pm$0.95 &4.21$\pm$0.30  \\
&Shen-368 &86.17$\pm$1.27 &87.56$\pm$3.24 &84.62$\pm$3.61 &93.73$\pm$0.87 &4.33$\pm$0.24  \\
\multirow{-4}{*}{BNT} &HCP-360 &85.74$\pm$1.06 &87.08$\pm$2.40 &84.38$\pm$2.77 &93.15$\pm$0.91 &4.43$\pm$0.26  \\
\cmidrule(lr){1-2} \cmidrule(lr){3-6} \cmidrule(lr){7-7}
\multicolumn{2}{c}{Atlas-free BNT (AC)} &{\bfseries 89.20$\pm$1.09} &{\bfseries 89.87$\pm$1.51} &{\bfseries 88.48$\pm$2.15} &{\bfseries 95.90$\pm$0.79} &{\bfseries 4.03$\pm$0.22}  \\
\multicolumn{2}{c}{Atlas-free BNT (SC)} &88.84$\pm$1.18 &89.74$\pm$2.30 &87.84$\pm$2.26 &95.70$\pm$0.08 &4.06$\pm$0.21  \\
\bottomrule
\end{tabular}
}
\end{table}

Resting-state fMRI reveals intrinsic functional-connectivity patterns that systematically differ between males and females. This allows us to train a machine learning model to predict the biological sex from these connectivity features. For this classification task, elastic net adopts the logistic regression formulation for training, all other classification models were trained by minimizing the cross-entropy loss. The classification accuracy, sensitivity, specificity and the AUROC obtained from each approach are given in Table \ref{tab:main_results}. The proposed atlas-free BNT consistently outperforms all atlas-based baselines. Within the atlas-free framework, we observe that parcellations derived using agglomerative clustering (AC) yield slightly better performance than those produced by spectral clustering (SC). This advantage likely arises because AC is more effective at generating internally homogeneous clusters. However, AC operates in a stepwise, greedy fashion, merging two clusters at a time, and therefore requires considerably more computation time than SC. Across the atlas-based approaches, connectivity-driven atlases substantially outperform the anatomical AAL atlas, with the Craddock-400 atlas achieving the best overall results.

\subsection{Brain Connectome Age Prediction}

\begin{figure*}[tbp]
\centering
\subfigure[Atlas-free BNT (AC)]{
\label{fig:atlas_free_bnt_ac_age}
\includegraphics[width=0.45\textwidth]{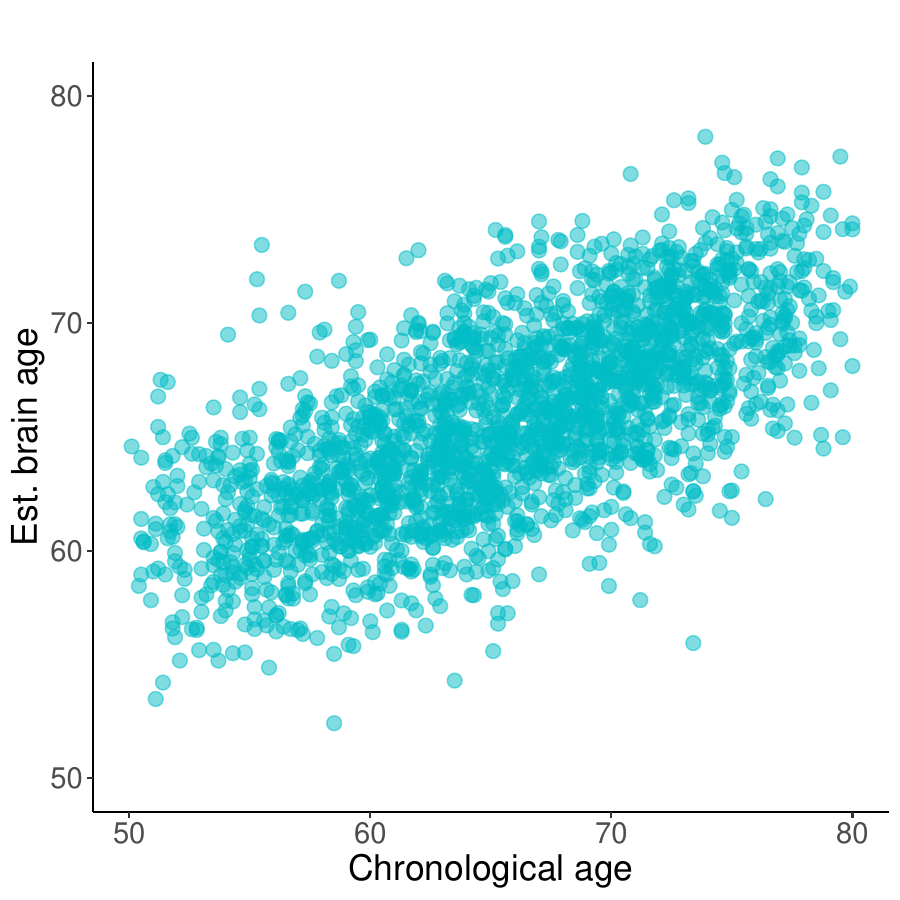}}
\subfigure[Atlas-free BNT (SC)]{
\label{fig:atlas_free_bnt_sc_age}
\includegraphics[width=0.45\textwidth]{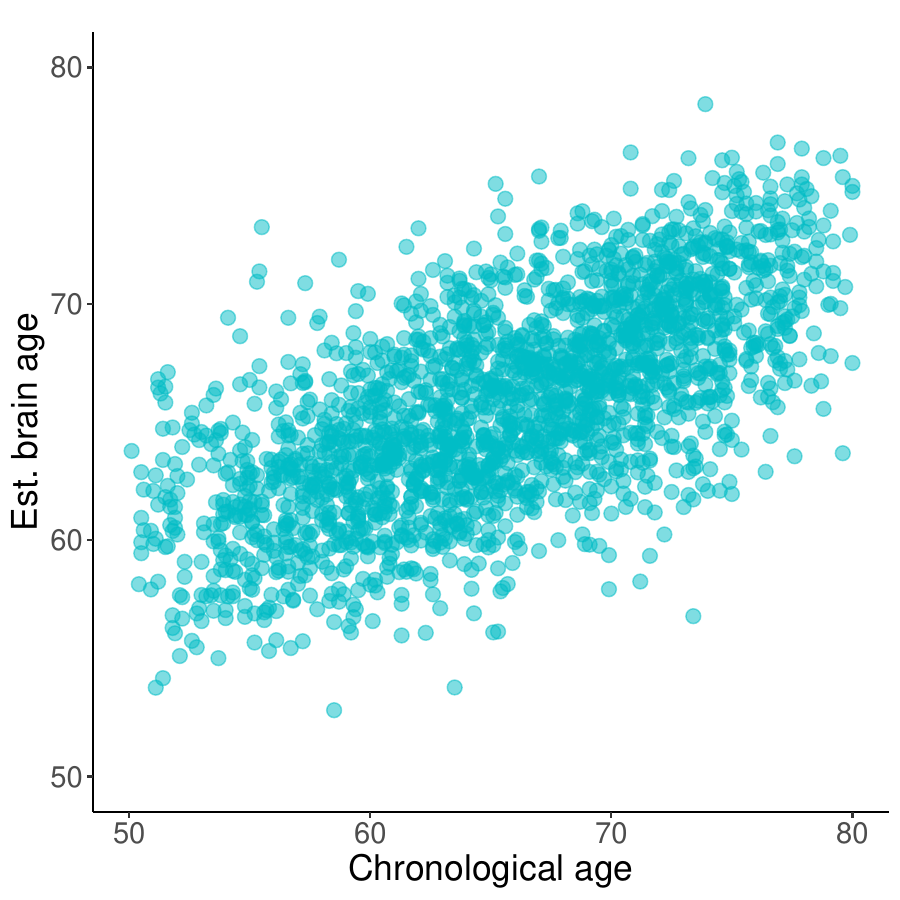}}
\caption{Atlas-free BNT's predicted brain-connectome age versus chronological age under two individualized brain parcellations: a) agglomerative clustering (AC) and b) spectral clustering (SC). Each point represents one subject/session. The mean and standard deviation of the absolute errors are calculated for both methods: a) $4.03\pm2.99$ years; b) $4.06\pm3.03$ years.}
\label{fig:atlas_free_bnt_age}
\end{figure*}

Brain-connectome age is an estimate of how old a person's brain appears based on neurobiological connectivity features extracted from structural or functional neuroimaging data. It may differ from chronological age, reflecting individual variability in the brain's aging process. The estimation model is typically trained on data from a cohort of healthy individuals, allowing brain-connectome age to serve as a benchmark for normative brain development. A higher or lower brain-connectome age relative to chronological age indicates accelerated or decelerated brain aging, respectively. This measure has been widely used as a biomarker to assess brain health and to aid in the early detection of neurodegenerative diseases. For this regression task, elastic net follows a standard linear regression formulation, all other regression models were trained by minimizing the mean squared error (MSE) between the brain-connectome age and chronological age. For evaluation, we report mean absolute error (MAE), which is commonly used in brain-age studies because it provides a more intuitive measure of prediction error in years than MSE. The obtained MAEs from different approaches are also given in Table \ref{tab:main_results}. The atlas-free BNT consistently outperforms atlas-based methods, demonstrating greater robustness and generalizability. Figure~\ref{fig:atlas_free_bnt_age} shows the atlas-free BNT’s predicted brain-connectome age plotted against chronological age, and the figures for atlas-based approaches are presented in Appendix~\ref{app:atls_brain_age_prediction}. Among atlas-based approaches, connectivity-driven atlases generally achieve better performance than the anatomical AAL atlas, with the exception of BrainGNN. Across all atlas-based baselines, the Craddock-400 atlas remains the strongest performer.

\subsection{Ablation Study}

\begin{table}[tbp]
\caption{Ablation study for the proposed atlas-free BNT (AC).}
\vspace{1em}
\label{tab:ablation_study}
\centering
\resizebox{\columnwidth}{!}{
\begin{tabular}{llcccccccccccccccc}
\toprule
& &\multicolumn{2}{c}{BNT} &\multicolumn{4}{c}{\textbf{Sex classification}} &\textbf{Brain age prediction}\\ \cmidrule(lr){3-4} \cmidrule(lr){5-8} \cmidrule(lr){9-9}
\multirow{-2}{*}{Method} &\multirow{-2}{*}{FNN} &MHSA &OCR &Accuracy &Sensitivity &Specificity &AUROC & Mean Absolute Error  \\ \cmidrule(lr){1-4} \cmidrule(lr){5-8} \cmidrule(lr){9-9}
V1 &\XSolidBrush  &\Checkmark &\Checkmark &88.53$\pm$1.32 &88.98$\pm$1.85 &88.03$\pm$1.85 &95.18$\pm$0.81
 &4.28$\pm$0.20\\
V2 &\Checkmark &\XSolidBrush  &\XSolidBrush &88.54$\pm$1.25 &89.33$\pm$1.97 &87.70$\pm$1.97 &95.23$\pm$0.80
 &4.95$\pm$0.27\\
V3 &\Checkmark &\Checkmark &\XSolidBrush &88.78$\pm$0.92 &89.36$\pm$1.49 &88.16$\pm$1.49 &95.34$\pm$0.70
 &\textbf{4.05$\pm$0.21}\\
V4 &\Checkmark &\XSolidBrush &\Checkmark &88.50$\pm$1.35 &89.40$\pm$2.07 &87.56$\pm$2.06 &95.16$\pm$0.84 &4.20$\pm$0.22\\ \cmidrule(lr){1-4} \cmidrule(lr){5-8} \cmidrule(lr){9-9}
Proposed &\Checkmark &\Checkmark &\Checkmark &\textbf{89.20$\pm$1.09} &\textbf{89.87$\pm$1.51} &\textbf{88.48$\pm$2.15} &\textbf{95.90$\pm$0.79} &4.06$\pm$0.21\\
\bottomrule
\end{tabular}
}
\end{table}

We conduct an ablation study to assess how individual components of the atlas-free BNT contribute to predictive performance. Specifically, we choose agglomerative clustering (AC) for brain parcellation and evaluate the following model variants, each obtained by removing or simplifying a single module while keeping the remaining pipeline unchanged:

\begin{itemize}
\item \textbf{V1 (No FNN feature transform)}: In the full model, the feedforward neural network (FNN) maps the connectivity feature vector $\vg$ to a latent representation $\vq$ that is better suited for subsequent block-wise node feature construction. In this variant, we remove the FNN and form node features by directly sum-pooling $\vg$ within each spatial block.

\item \textbf{V2 (No BNT; MLP on pooled tokens)}: The BNT module comprises multi-head self-attention (MHSA) and the orthonormal clustering readout (OCR), it models interactions among block tokens and aggregates them into a subject-level embedding. In this variant, we remove the entire BNT module and instead concatenate all block-level node features and feed the resulting vector directly into the downstream MLP regressor.

\item \textbf{V3 (No OCR readout)}: In the full model, OCR summarizes MHSA-updated node features via soft clustering to produce the final graph-level embedding. In this variant, we remove OCR and replace it with a standard MHSA-only aggregation for prediction (i.e., the transformer encoder is retained, but the OCR-based readout is omitted).

\item \textbf{V4 (No MHSA transformer)}: In the full model, the MHSA transformer captures the long-range dependencies and inter-block interactions, it infuses global information into the node representations before they are clustered by OCR. In this variant, we remove the MHSA encoder and apply OCR directly to the original block-level node features for downstream prediction.
\end{itemize}

Across both tasks, the ablation results in Table \ref{tab:ablation_study} suggest that different components contribute in complementary ways, with the overall framework being most beneficial when all modules are used together. For sex classification, performance is relatively stable across the four variants, and the full atlas-free BNT remains the top performer. This pattern indicates that sex-related features are likely strong and broadly distributed in the connectivity representation, so even simplified aggregation strategies can recover much of the discriminative information; nevertheless, retaining the full pipeline provides a consistent edge, likely by better standardizing features and reducing sensitivity to subject-specific parcellation variability. 

In contrast, brain-age prediction shows clear separation among the four variants, implying that age-related effects are more subtle and depend more heavily on how information is integrated across the brain. In particular, the full model achieves an MAE of ($4.06\pm0.21$) years. Removing the BNT module entirely and feeding concatenated block features directly to an MLP (V2) yields the largest performance drop ($4.95\pm0.27$). This shows that modeling inter-block relationships and performing structured global aggregation via BNT are critical for accurate age prediction. In contrast, removing the OCR readout while retaining MHSA (V3) produces essentially no degradation ($4.05\pm0.21$). It suggests that for this task the transformer encoder alone already learns a sufficiently informative global representation and that OCR provides only marginal additional benefit. Eliminating the MHSA encoder and applying OCR directly to the uncontextualized block features (V4) increases error to $4.20\pm0.22$, which demonstrates the key role of self-attention in encoding each block embedding with whole-brain context before pooling. Finally, bypassing the FNN feature transform and pooling the original connectivity vectors within blocks (V1) also degrades performance ($4.28\pm0.20$), indicating that the learned non-linear mapping helps produce block-level features that are more suitable for subsequent feature aggregation. Overall, these results reinforce that the atlas-free BNT’s main advantage (especially for brain-age prediction) comes from transformer-based global context modeling, while the FNN and OCR contribute additional gains that are smaller but complementary.

\subsection{Discussion}

\begin{figure*}[tbp]
\centering
\subfigure{
\includegraphics[width=0.45\textwidth]{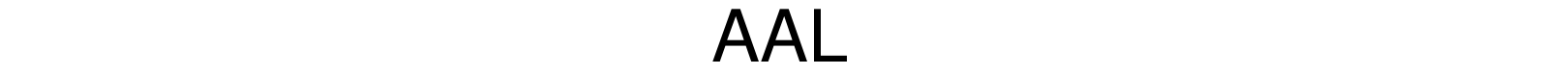}}
\subfigure{
\includegraphics[width=0.45\textwidth]{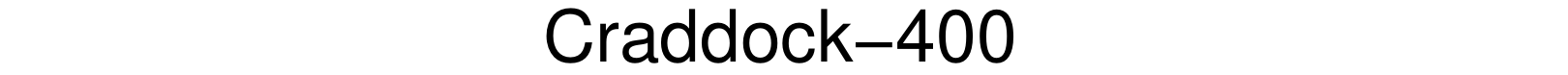}}\\
\subfigure{
\includegraphics[width=0.45\textwidth]{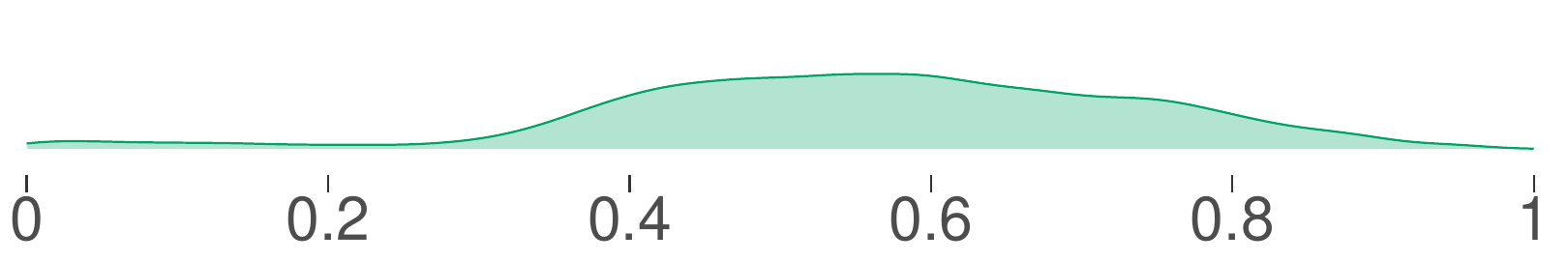}}
\subfigure{
\includegraphics[width=0.45\textwidth]{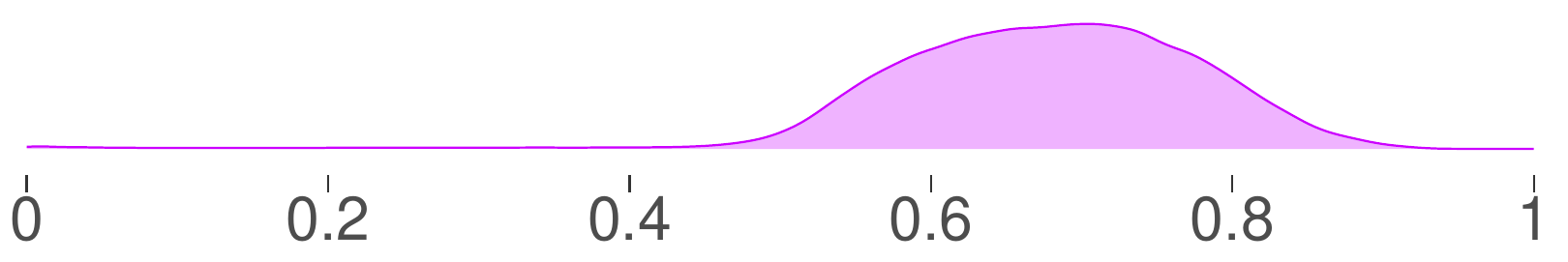}}\\
\subfigure{
\includegraphics[width=0.45\textwidth]{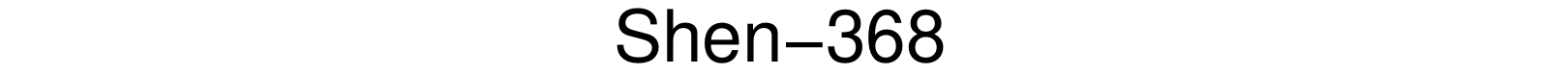}}
\subfigure{
\includegraphics[width=0.45\textwidth]{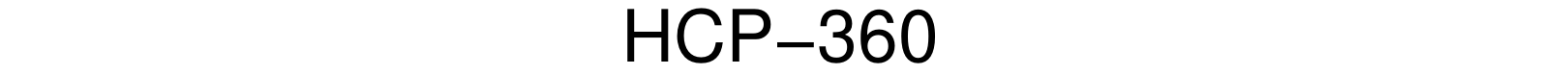}}\\
\subfigure{
\includegraphics[width=0.45\textwidth]{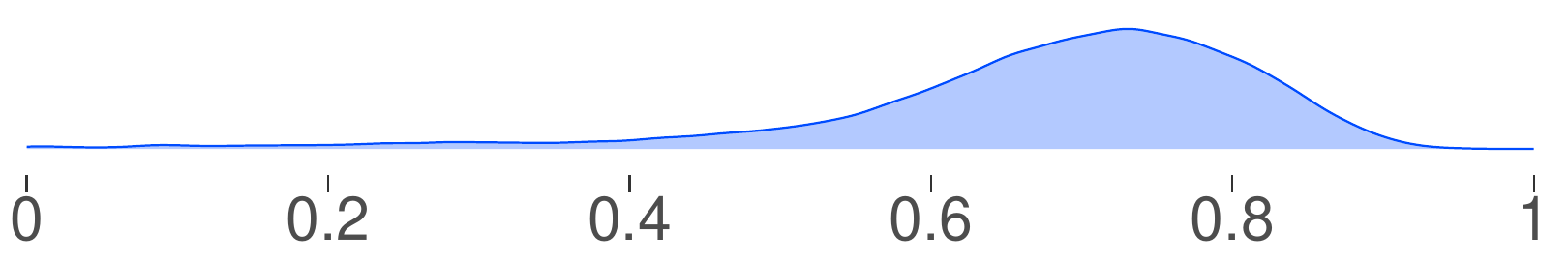}}
\subfigure{
\includegraphics[width=0.45\textwidth]{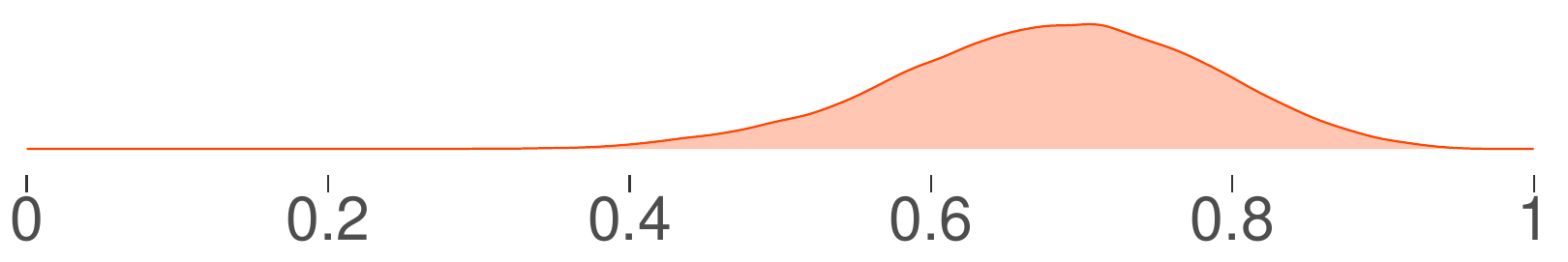}}\\
\subfigure{
\includegraphics[width=0.45\textwidth]{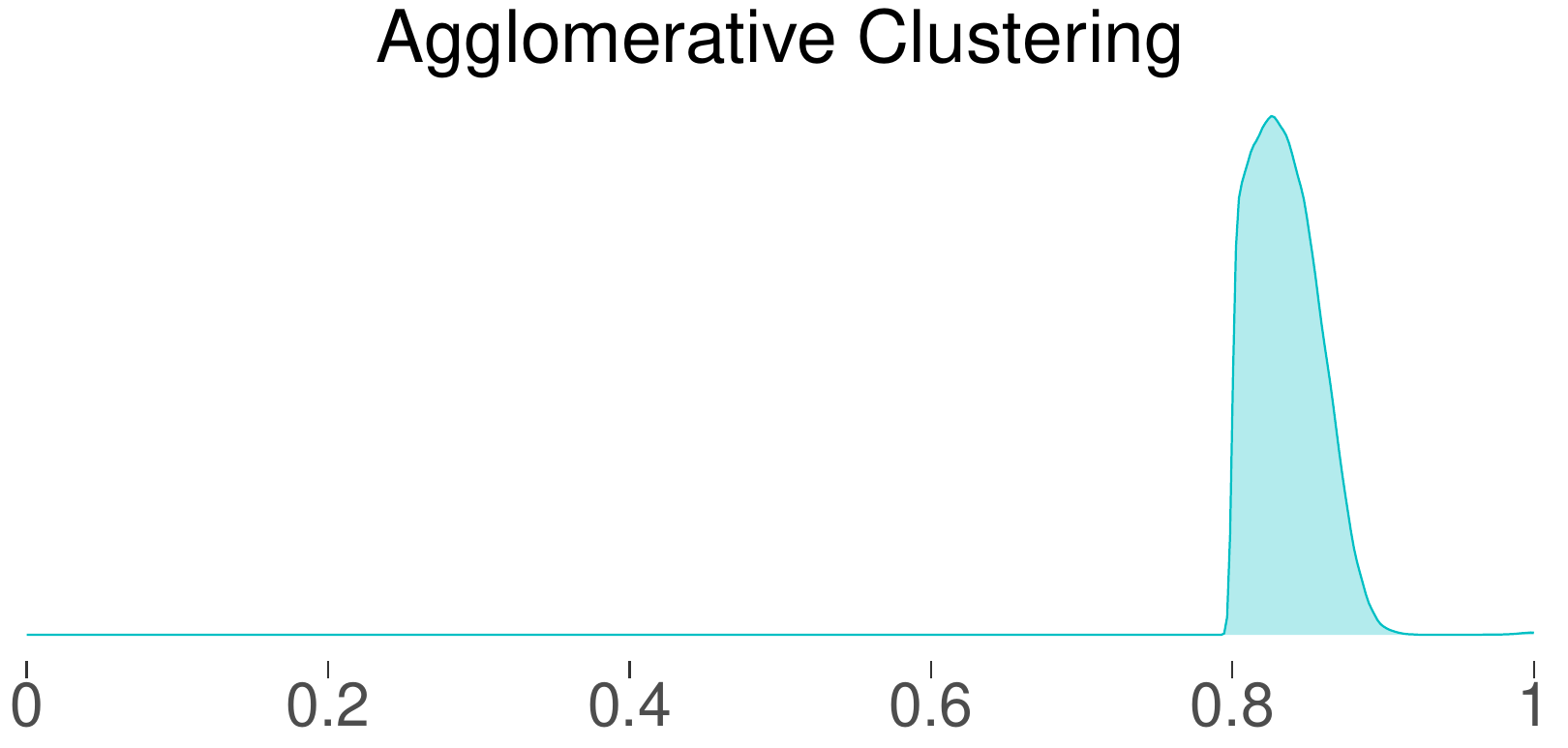}}
\subfigure{
\includegraphics[width=0.45\textwidth]{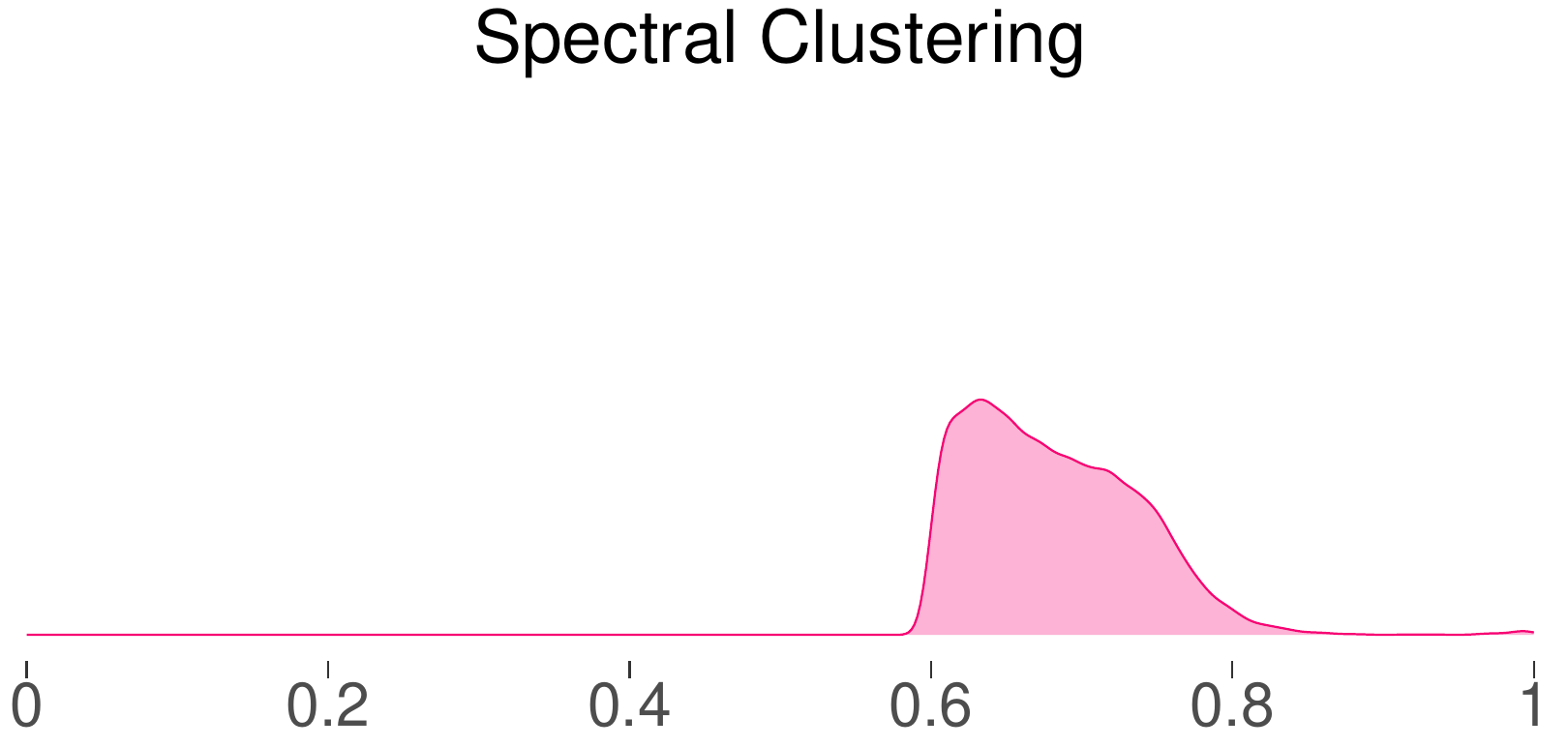}}
\caption{The distributions of averaged intra-correlations between voxel-wise BOLD time series and the corresponding mean time series within each ROI across 100 subjects. Atlas-based brain parcellation methods (AAL, Craddock-400, Shen-368, and HCP-360) lead to lower functional homogeneity compared to individualized brain parcellation methods (agglomerative and spectral clustering).}
\label{fig:roi_homogeneity}
\end{figure*}

The proposed atlas-free BNT leverages individualized brain parcellation to mitigate ROI misalignment and enhance functional homogeneity. Functional homogeneity is quantified by computing intra-correlations between voxel-wise BOLD time series and the corresponding mean time series within each ROI. Fig. \ref{fig:roi_homogeneity} presents the distributions of average intra-correlations across 100 subjects obtained using different parcellation methods. In atlas-based approaches, a fixed atlas is applied uniformly to all subjects, leading to reduced homogeneity due to inter-subject variability. By contrast, individualized parcellation adapts to each subject’s unique brain organization and allows homogeneity to be tuned by setting parameters such as the merging correlation threshold $\nu$ in agglomerative clustering (AC) or the correlation cutoff threshold $\tau$ and the number of clusters $k$ in spectral clustering (SC), resulting in consistently higher functional homogeneity. Notably, SC homogeneity can be further increased by using a larger $k$, which typically yields smaller ROIs with more homogeneous signals; in our experiments, we set $k=400$ to achieve the best overall predictive performance.

AC and SC produce individualized parcellations through fundamentally different mechanisms, which leads to distinct ROI configurations. AC is effectively deterministic given a fixed similarity measure and stopping rule, since the greedily merged sequence is uniquely determined. By comparison, SC introduces randomness through the $k$-means step used to cluster the spectral embeddings; different initializations can lead to different solutions, and occasional poor local optima may require rerunning $k$-means with a new initialization to obtain stable parcellations.

Another advantage of the atlas-free BNT framework lies in its robustness to residual misregistration arising during spatial normalization. Even with modern methods, inter-individual anatomical variability yields imperfect alignment to a common template (e.g., MNI), which can propagate bias into downstream brain network analysis. To mitigate this, atlas-free BNT partitions the template space into overlapping 3D blocks and aggregates local functional-connectivity features within each block. This design cushions small spatial shifts: signals that would otherwise be displaced are still pooled together within shared neighborhoods, and the transformer layers then standardize these block-level features in latent space. In effect, residual misregistration behaves like a soft, local perturbation rather than a hard boundary error, reducing sensitivity to inter-subject variability and improving the inference stability.

\begin{figure*}[tbp]
\centering
\subfigure[Sex classification]{
\label{fig:grad_cam_sex}
\includegraphics[width=\textwidth]{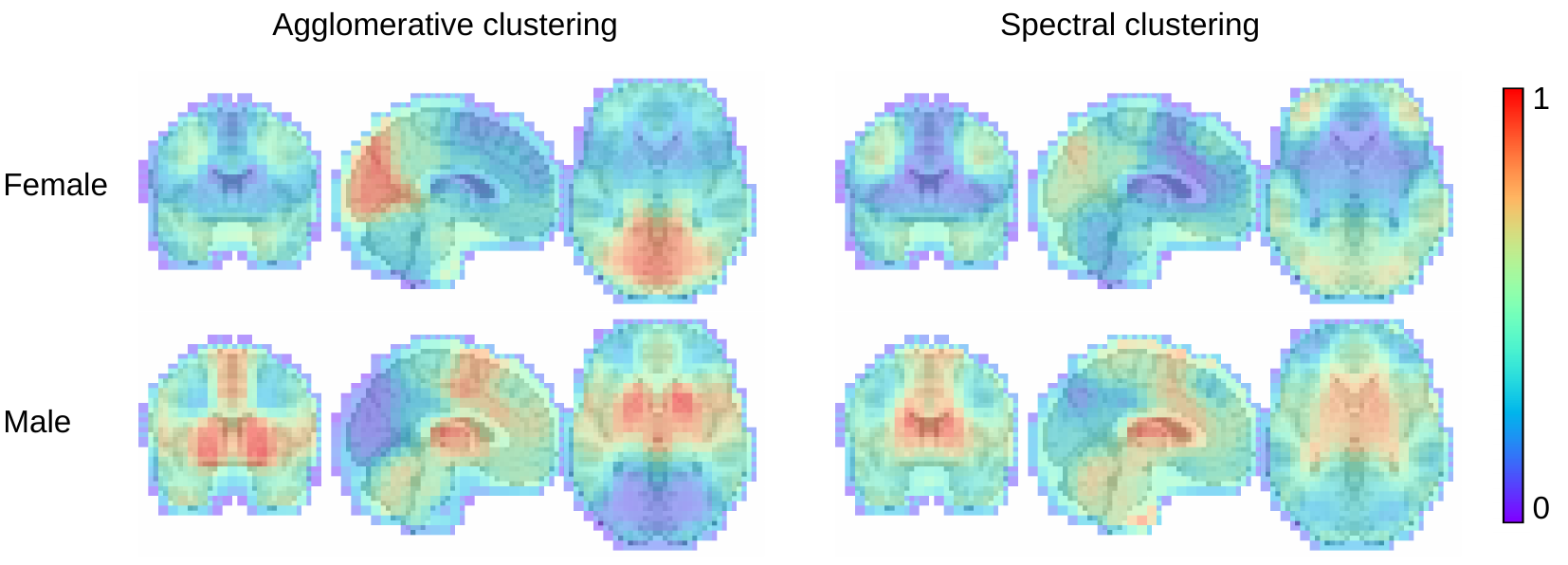}}\\
\subfigure[Brain-connectome age prediction]{
\label{fig:grad_cam_age}
\includegraphics[width=\textwidth]{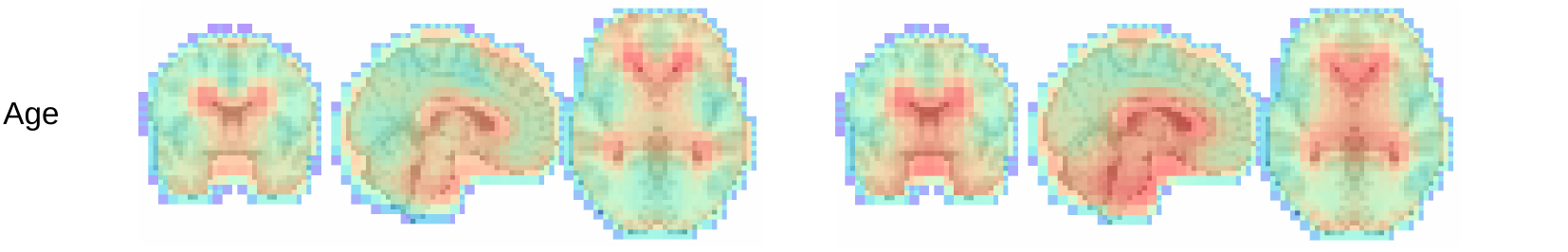}}
\caption{Averaged Grad-CAM saliency maps overlayed on the MNI brain template highlight the regional contributions to the atlas-free BNT model's inference. Warmer colors indicate regions with stronger positive influence on the output, revealing the brain areas most relevant to sex classification and brain-connectome age prediction.}
\label{fig:grad_cam}
\end{figure*}

For interpretability, we applied Grad-CAM \cite{selvaraju2017grad} to generate saliency maps highlighting regional contributions to the atlas-free BNT model’s inference. Subject-level maps were computed and averaged within one cross-validation fold (Fig.~\ref{fig:grad_cam}). In particular, the saliency maps in Fig. \ref{fig:grad_cam_sex} reveal distinct patterns of brain regions contributing to sex classification. For the female classification map, the most prominent saliency is observed in the posterior occipital cortex, suggesting that visual and posterior association networks provide key evidence for predicting the female class. In contrast, the male classification map shows stronger contributions in the frontal cortical regions, as well as pronounced saliency in subcortical structures including the thalamus and basal ganglia, together with the cerebellum. This indicates that male predictions rely on an integrated set of executive, subcortical, and cerebellar networks. Overall, the maps suggest that the model captures sex-differentiated contributions spanning both cortical and subcortical systems to achieve robust classification.

The Grad-CAM map for the brain-connectome age model in Fig.~\ref{fig:grad_cam_age} shows widespread, high saliency across the cortical gray-matter mantle (spanning frontal, parietal, temporal, and occipital cortices, including the medial wall) and equally strong saliency in subcortical nuclei (thalamus/striatal complexes) and the posterior cerebellum (vermis and hemispheric lobules). This pan-cortical plus subcortico-cerebellar pattern indicates that age estimates are derived from a distributed cortico–subcortical–cerebellar network rather than a focal locus, consistent with broad lifespan effects including global cortical thinning \cite{Frangou2022_HBM}, frontostriatal and thalamo-cortical connectivity reconfiguration with aging \cite{Webb2020_NeurobiolAging,Das2021_CerebCortex,Geerligs2015_CerebCortex}, and cerebellar structural change across adulthood \cite{Arleo2024_Cerebellum}. We emphasize that Grad-CAM highlights supportive evidence within this trained model (not causal mechanisms), and that block-wise reconstruction and cross-subject averaging may smooth boundaries; nevertheless, the convergence of strong cortical, subcortical, and cerebellar saliency supports a network-level basis for the model’s age predictions. It is also reassuring to observe that aggolerative and spectral clustering approaches provide similar salience maps for both sex and age prediction, suggesting these results are more likely to be from intrinsic biological effects rather than artifacts of the specific technical implementations employed.

\section{Conclusion}
In this paper, we proposed a novel atlas-free Brain Network Transformer (atlas-free BNT) framework to address inherent limitations associated with traditional atlas-based brain network analyses, including spatial misalignment, ROI heterogeneity, and atlas-selection bias. By leveraging individualized brain parcellations derived directly from subject-specific rs-fMRI data, our approach accurately captures unique functional connectivity patterns for each participant. To ensure cross-subject comparability, we introduced a standardized voxel-based connectivity representation processed through a transformer architecture. Extensive experiments on sex classification and brain-connectome age prediction demonstrated that our atlas-free BNT method achieves superior performance and generalizability compared to multiple state-of-the-art atlas-based methods.


We also acknowledge several limitations of the proposed atlas-free BNT and plan to address them in future work. Although the method is designed to be robust to imperfect alignment, it still relies on spatial normalization to a common template in the MNI space; we will therefore investigate strategies that further reduce sensitivity to misregistration, including improved alignment quality control and more deformation-tolerant feature aggregation. Because individualized parcellation depends on clustering hyperparameters (e.g., $\nu$, $\tau$ and $k$), we will study principled selection criteria and stability analyses to reduce sensitivity to these choices and improve reproducibility across datasets/tasks. We also plan to optimize the computational pipeline to facilitate deployment on larger cohorts. Finally, we will conduct broader multi-site and clinical validations, and explore domain-adaptation strategies, to ensure that atlas-free BNT generalizes reliably across acquisition settings and diverse patient populations.

\section{Acknowledgement}
The study was supported by the National Institutes of Health grants/awards (R01AG089806, R01AG070937, R01AG072603, and P30AG066511).

\bibliographystyle{myunsrt}
\bibliography{Bibliography}

\begin{appendices}
\section{Parameter Tuning}
\label{app:hyperparameter}
To ensure a fair comparison against the four atlas-based baselines, we tuned their hyperparameters following the procedures summarized below:
\begin{itemize}
\item For the elastic net, two regularization hyperparameters, $\lambda_1$ and $\lambda_2$, control the strengths of the $l_1$- and $l_2$-norm penalties, respectively. We set $\lambda_1=1e^{-5}$ and $\lambda_2=5e^{-5}$ for the sex-classification task, and $\lambda_1 = 0.04$ and $\lambda_2=0$ for brain-connectome age prediction.
\item For BrainGNN, we performed a grid search over learning rate $\{0.001,0.005,0.01\}$, feature dimension $\{100,200,300,500\}$, and number of ROIs $\{50,100,200,300\}$; the best validation configuration was learning rate $0.005$, feature dimension $200$, and $200$ ROIs.
\item For Graphormer, we set the learning rate to $1e^{-4}$ and used two encoder layers, four attention heads, and 256-dimensional embeddings.
\item For the original BNT, we adopted the default cosine-annealing learning-rate scheduling strategy that starts with $1e^{-4}$, and used two encoder layers, four attention heads, and 360-dimensional embeddings.
\end{itemize}

\section{Brain Connectome Age Prediction}
\label{app:atls_brain_age_prediction}
\begin{figure*}[tbp]
\centering
\subfigure[Elastic net (Craddock-400)]{
\label{fig:elastic_net_age}
\includegraphics[width=0.45\textwidth]{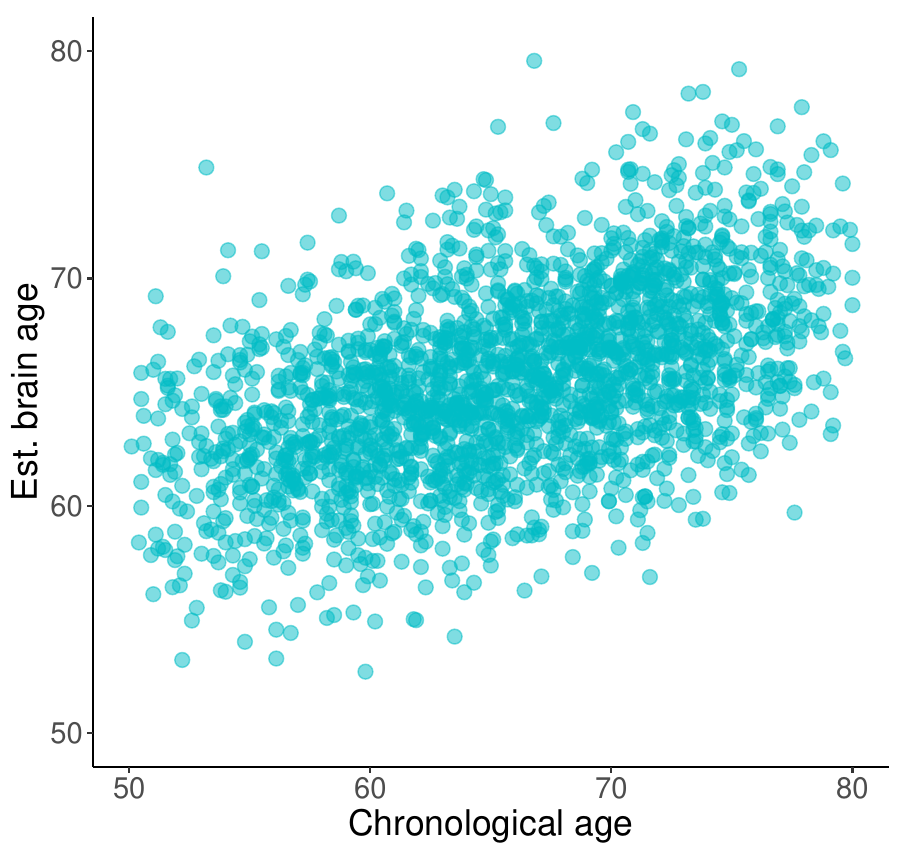}}
\subfigure[BrainGNN (AAL)]{
\label{fig:brain_gnn_age}
\includegraphics[width=0.45\textwidth]{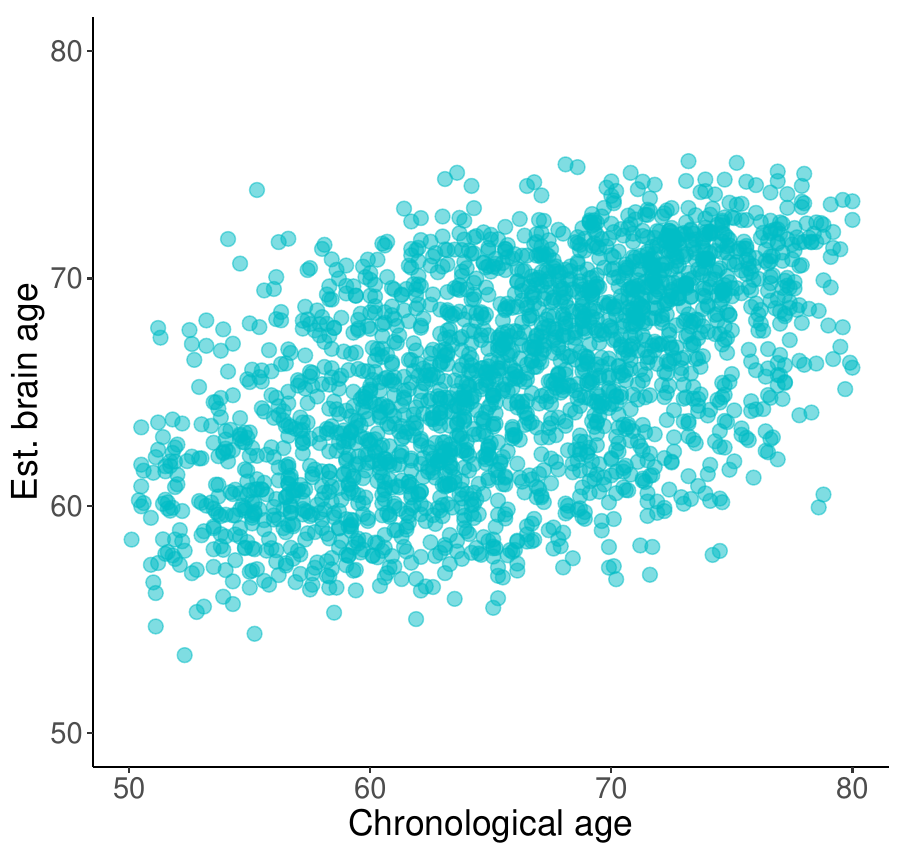}}\\
\subfigure[Graphormer (Craddock-400)]{
\label{fig:graphormer_age}
\includegraphics[width=0.45\textwidth]{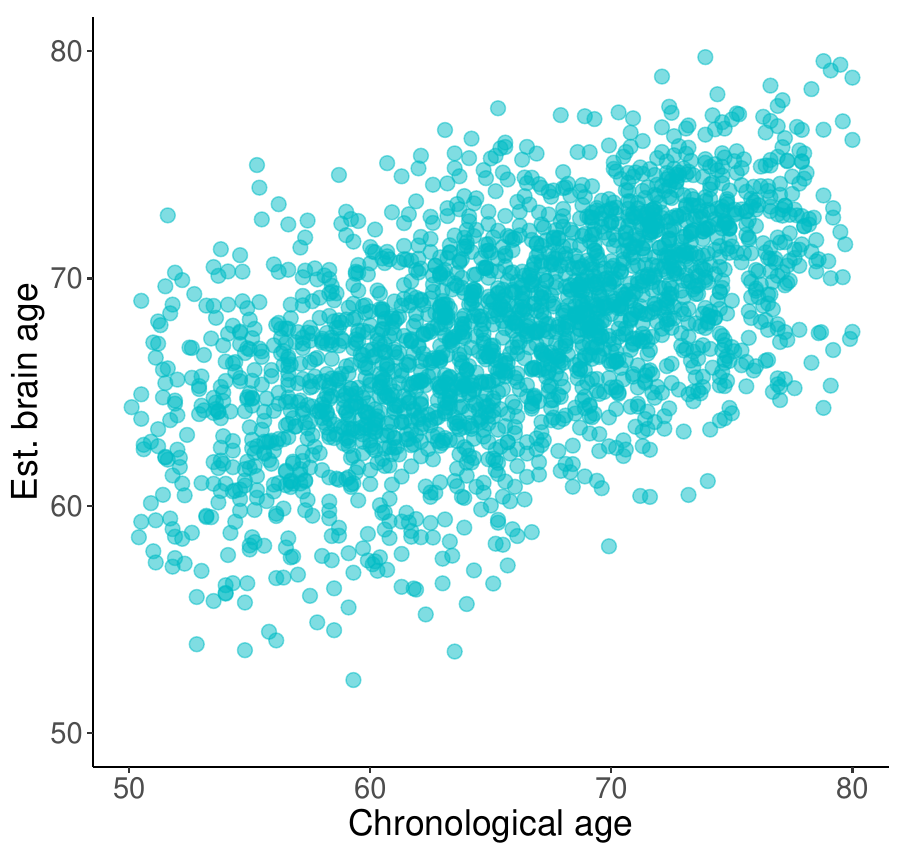}}
\subfigure[BNT (Craddock-400)]{
\label{fig:bnt_age}
\includegraphics[width=0.45\textwidth]{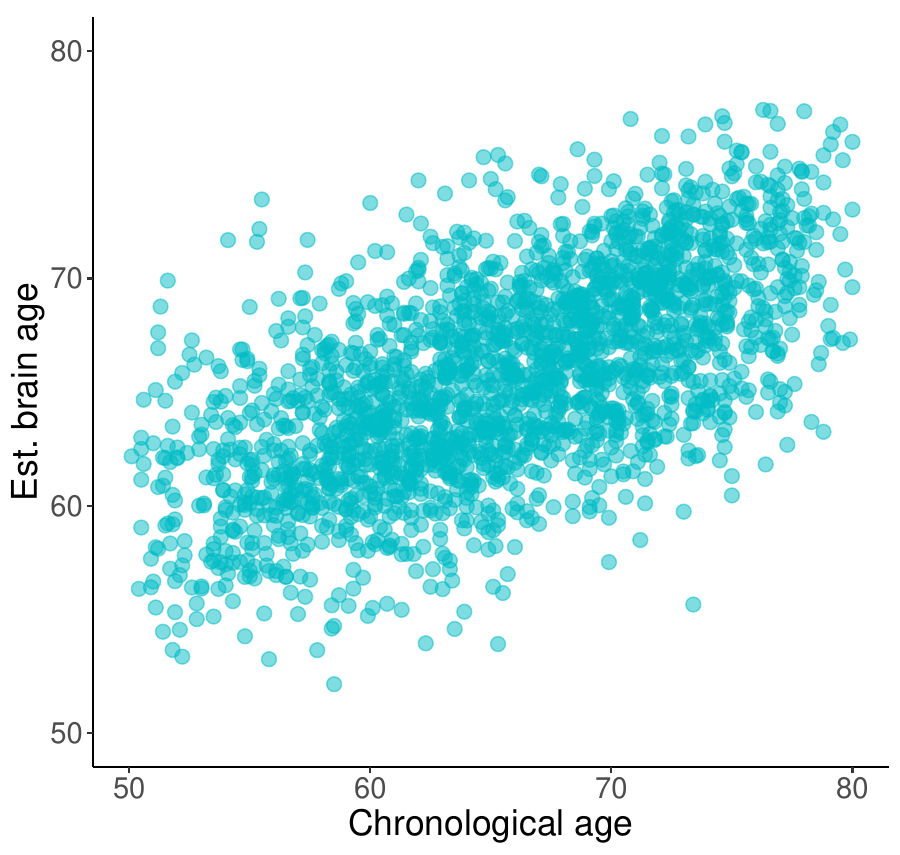}}
\caption{The predicted brain-connectome age versus chronological age by the atlas-based approaches with their best performing atlases: a) Elastic net (Craddock-400), b) BrainGNN (AAL), c) Graphormer (Craddock-400), d) BNT (Craddock-400). Each point represents one subject/session. The mean and standard deviation of the absolute errors are calculated for all methods: a) $4.86\pm3.54$ years; b) $4.63\pm3.47$ years; c) $4.71\pm3.64$; d) $4.21\pm3.21$.}
\label{fig:atlas_based_age}
\end{figure*}

Fig. \ref{fig:atlas_based_age} shows the four atlas-based approaches’ predicted brain-connectome age plotted against chronological age.

\end{appendices}

\end{document}